\documentclass[useAMS,usegraphicx]{mn2e}
\usepackage{rotate}
\usepackage{rotating}
\usepackage{times}
\newif\ifAMStwofonts
\AMStwofontstrue
\def\gs{\mathrel{\hbox{\rlap{\hbox{\lower4pt\hbox{$\sim$}}}\hbox{$>$}}}}
\def\ls{\mathrel{\hbox{\rlap{\hbox{\lower4pt\hbox{$\sim$}}}\hbox{$<$}}}}
\def\Msunpyr{M$_{\odot}\,$yr$^{-1}$}
\def\chandra{{\it Chandra}}
\def\suzaku{{\it Suzaku}}

\def\xmm{{\it XMM-Newton}}

\def\astroh{{\it Astro-H}}

\def\et{{et al.\ }}

\def\mrk79{{Mrk~79}}
\def\pg1211{{PG~1211+143}}

\def\3c{{3C~273}}
\def\rg{{\thinspace r_{\rm g}}}

\def\redchi{{\chi^2_\nu}}

\def\feka{{Fe~K$\alpha$}}

\def\fexxv{{Fe~\textsc{xxv}}}
\def\fexxvi{{Fe~\textsc{xxvi}}}
\def\nh{{N_{\rm H}}}

\def\deg{^{\circ}}

\def\cm{{\rm\thinspace cm}}
\def\erg{{\rm\thinspace erg}}
\def\eV{{\rm\thinspace eV}}

\def\ga{{\rm\thinspace gauss}}

\def\keV{{\rm\thinspace keV}}
\def\km{{\rm\thinspace km}}

\def\Msun{\hbox{$\rm\thinspace M_{\odot}$}}

\def\s{{\rm\thinspace s}}

\def\ps{{\rm\thinspace s^{-1}}}

\def\yr{{\rm\thinspace yr}}

\def\ergpscmps{\hbox{$\erg\cm^{-2}\s^{-1}\,$}}

\def\ergcmps{\hbox{$\erg\cm\s^{-1}\,$}}

\def\kmps{\hbox{$\km\ps\,$}}

\def\Msunpyr{\hbox{$\Msun\yr^{-1}\,$}}

\def\pscm{\hbox{$\cm^{-2}\,$}}

\title[\pg1211: Outflow or disc?]
      {
The origin of blue-shifted absorption features in the X-ray spectrum of \pg1211: Outflow or disc?
      }

\author[L. C. Gallo \& A. C. Fabian]
       {L. C. Gallo$^1$ 
and	A. C. Fabian$^2$ 
        \\ 
$^{1}$ Department of Astronomy and Physics, Saint Mary's University, 923 Robie Street, Halifax, NS, B3H 3C3, Canada \\
$^{2}$ Institute of Astronomy, University of Cambridge, Madingley Road, Cambridge CB3 0HA\\
}
\date{Accepted. Received. }
\pagerange{\pageref{firstpage}--\pageref{lastpage}}
\pubyear{2011}
\begin{document}
\maketitle
\label{firstpage}

\begin{abstract}
In some  radio-quiet active galaxies (AGN), high-energy absorption features in the x-ray spectra have been interpreted as Ultrafast Outflows (UFOs) -- highly ionised material (e.g. \fexxv\ and \fexxvi) ejected at mildly relativistic velocities.  In some cases, these  outflows can carry energy in excess of the binding energy of the host galaxy.  Needless to say, these features demand our attention as they are strong signatures of AGN feedback and will influence galaxy evolution.  For the same reason, alternative models need to be discussed and refuted or confirmed.  Gallo \& Fabian proposed that some of these features could arise from resonance absorption of the reflected spectrum in a layer of ionised material located above and corotating with the accretion disc.  Therefore, the absorbing medium would be subjected to similar blurring effects as seen in the disc.  A priori, the existence of such plasma above the disc is as plausible as a fast wind.  In this work, we highlight the ambiguity by demonstrating that the absorption model can describe the $\sim7.6\keV$ absorption feature (and possibly other features) in the quasar \pg1211, an AGN that is often described as a classic example of an UFO.  In this model, the $2-10\keV$ spectrum would be largely reflection dominated (as opposed to power law dominated in the wind models) and the resonance absorption would be originating in a layer between about $6$ and $60$ gravitational radii.  The studies of such features constitutes a cornerstone for future X-ray observatories like \astroh\ and {\it Athena+}.   Should our model prove correct, or at least important in some cases, then absorption will provide another diagnostic tool with which to probe the inner accretion flow with future missions.
 
\end{abstract}

\begin{keywords}
accretion, accretion discs --
black hole physics --
relativistic processes --
line: formation, identification --
galaxies: active --
X-ray: galaxies 
\end{keywords}

% --------------------------------------------------------------------------

\section{Introduction}
\label{sect:intro}

Blueshifted absorption features in the X-ray spectra of some radio-quiet active galactic nuclei (AGNs) have been attributed to 
resonant K-shell absorption lines of iron (\fexxv\ and \fexxvi) that are seen in outflow (e.g. Nandra \et 1999, 2007; Turner \et 2002; Dadina \et 2005; Longinotti \et 2007).  Many of these features are transient, seen only at
one epoch or during some part of an observation, and their existence is debated on statistical grounds (Vaughan \& Uttley 2008).  However,
other detections are more robust and have been observed on multiple occasions and/or with different instruments (e.g. Pounds \et 2003; Reeves \et 2009).  

To be associated with \fexxv\ or \fexxvi\ some of these features have to originate in very high velocity winds (e.g. $v\gs 10^4 \kmps$) or
even moving at mildly relativistic velocities ($v\gs 0.1c$) (e.g. Reeves \et 2009; Tombesi \et 2010; see Cappi 2006 for a review).  The claimed detection of such ultra-fast outflows (UFOs)  are widespread.  Tombesi \et (2010) suggest that as much as $\gs 35$ per cent of radio-quiet AGN may contain a UFO.  These UFOs can be very important in AGN feedback and thereby for understanding galaxy formation and the origin of the M-$\sigma$ relation (Kormendy \& Gebhardt 2001; Merritt \& Ferrarese 2001).  Therefore, alternatives to UFOs need to be discussed and refuted (or confirmed).

Gallo \& Fabian (2011; hereafter GF11) propose that some blueshifted absorption features could arise in a plasma that is  located above and co-rotating with the inner accretion disc (Ruszkowski \& Fabian 2000) and is optically thick at the energies of the resonance lines of iron.   Depending on the inclination and geometry of the plasma, narrow absorption features attributed to \fexxv\ and \fexxvi\ could be imprinted anywhere between about 4 and $10\keV$ (GF11).  Presumably, some features could be seen at even higher energies if originating from other transitions like  \fexxvi\ Ly$\beta$, with a rest energy at $E=8.25\keV$.
The appeal of this model is that the plasma is subjected to velocities that are already present in the disc, and does not require an additional launching mechanism.   This is not to suggest that disc winds are not present, but simply that the diversity of such x-ray features need to be better understood.

In this work we examine the quasar \pg1211\ to illustrate the ambiguity.  \pg1211\ is a classic example of an UFO candidate.  \xmm\ EPIC observations
in 2001 revealed a significant absorption feature at a rest frame energy of $\sim7.6\keV$ (Pounds \et 2003).  Later observations in 2004 and 2007 found a consistent feature at slightly lower significance (Pounds \& Reeves 2009; Reeves \& Pounds 2012).   The line was interpreted as blueshifted He- or H-like iron with an outflow velocity of either $\sim0.13c$ or $\sim0.08c$, respectively (Pounds \et 2003).  There were indications of other absorption features in the 2001
data between $1-3\keV$ that were attributed to ionised species of Ne, Mg, and S, also originating in a high velocity outflow.  These low-energy
features were not reported in the later \xmm\ observations nor the \chandra\ LETG observation (Reeves et al. 2005).   The iron feature may also be variable as it was not detected in a 2005 \suzaku\ observation (Gofford \et 2013; but see Patrick \et 2012).  The high velocity of the outflow was questioned by Kaspi \& Behar (2006) who argued that a much lower velocity ($\approx 3000\kmps$) was sufficient based ion-by-ion modeling of 
the \xmm\ RGS spectra.  However, the authors did acknowledge the poor signal-to-noise of their spectra and did not rule out a much higher velocity.

Most recently Tombesi \et (2011) constructed photoionisation models using {\sc xstar} (Kallman \& Bautista 2001) to describe a number of UFO candidates.  Specifically with respect to \pg1211\ they modeled the 2001 EPIC-pn
spectrum between $4-10\keV$ with a highly ionised absorber with an ionisation parameter\footnote{The ionisation parameter is $\xi = L/nr^2$ where $L$ is the ionising luminosity between $1-1000$ Ryd, $n$ is the number density of electrons, and r is the distance of the gas from the ionising source.} of log$\xi \approx 2.87$ and a column
density of $\nh \approx 8 \times 10^{22}\pscm$ outflowing at $v\approx0.15c$.  The model was consistent with that produced by Pounds \& Page (2006).  Adopting a mass of $\sim10^8 \Msun$ for \pg1211\ results in an outflow mass rate of $\sim 5\Msunpyr$ (Pounds \& Page 2006; Reeves \& Pounds 2012).
Reeves \& Pounds (2012) further demonstrate how the kinetic power is a significant fraction of the quasar bolometric luminosity ($> 10$ per cent), and can be a
strong source of mechanical feedback in the host galaxy.

In this work we present an alternative to the outflow model for \pg1211.  Namely, we describe the absorption feature as arising from resonance iron absorption
close to the black hole where the plasma is subject to blurring effects (GF11).

\section{A possible model for \pg1211}
\label{sect:picture}

We modeled the 2001 \xmm\ EPIC-pn spectrum of \pg1211\ between $2-10\keV$ in order to focus on the significant absorption feature seen at about
$7\keV$ ($\sim 7.6\keV$ in the AGN rest frame).  The energy and width of the feature during other \xmm\ observations are reported as being comparable to the 2001 measurements, but detected at lower significance. 
The works discussed above mainly fit this spectral region with an absorbed power law and, if necessary, a narrow, neutral \feka\ emission line.
Here, we fit the continuum with a blurred reflection plus power law model (Ross \& Fabian 2005).  As with other works, for the time being, we ignore the spectrum below $2\keV$ where complications arise due to the soft-excess and distant ionised absorption/emission.

% --------------------------------------------------------------------------
\begin{figure}
\rotatebox{270}
{\scalebox{0.32}{\includegraphics{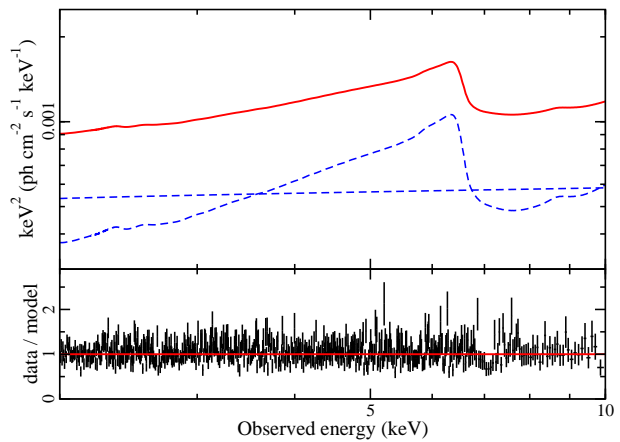}}}
\caption{The $2-10\keV$ spectrum of \pg1211\ is described with a blurred reflection plus power law model.  
Upper panel:  The model components (i.e. the power law and
blurred reflector) are shown by the blue, dashed curves.  The combined model is shown by the solid, red curve.  Lower panel:  The residuals (data/model)
remaining when the model is fitted to the data.  The model describes the spectrum well but does not account for the deviations at about $7\keV$.
}
\label{fig:ref}
\end{figure}
% --------------------------------------------------------------------------
The model describes the spectrum rather well ($\redchi=1.03$; Fig.~\ref{fig:ref}).  The power law continuum has a photon index of $\Gamma\approx 1.9$ and the ionisation parameter of the disc is $\xi\approx 290\ergcmps$ with an iron over-abundance of $\sim5$ compared to solar.  The blurring
parameters are typical compared to other type 1 AGN (e.g. Crummy \et 2006; Walton \et 2013).  The disc is inclined $i\approx35\deg$ and the inner edge of the disc extends down to $R_{in} \approx 1.5\rg$ ($1\rg = 1 GM/c^2$).  The emissivity profile of the disc is represented by a broken power law where the
inner profile is $q_{in}\approx 5$ and flattens to $q_{out}=3$ beyond about $6\rg$.  Between about $2-10\keV$ the spectrum is dominated by the
reflection component (Fig.~\ref{fig:ref}, top panel).  
Despite the ionisation parameter of the disc falling in the range where fluorescence is suppressed by resonant Auger destruction, the high iron abundance still allows production of a strong emission line and edge.
 The lower panel of Fig.~\ref{fig:ref} clearly show the residuals remaining around $7\keV$.

% --------------------------------------------------------------------------
\begin{figure*}
\begin{center}
\begin{minipage}{0.5\linewidth}
\scalebox{0.32}{\includegraphics[angle=0]{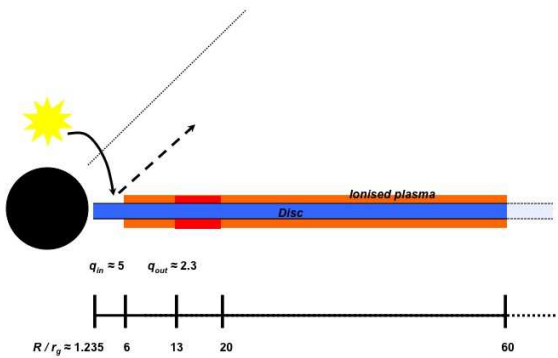}}
\end{minipage}  \hfill
\begin{minipage}{0.48\linewidth}
\scalebox{0.32}{\includegraphics[angle=270]{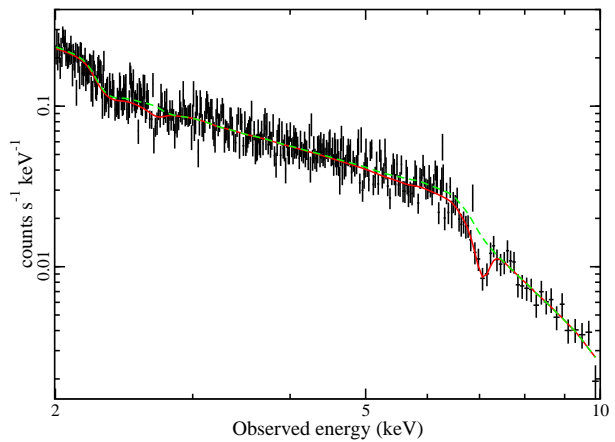}}
\end{minipage}
\end{center}
\caption{
Left panel:  A cartoon illustrating the potential geometry of the inner accretion disc that could reproduce the $2-10\keV$ spectrum (right panel).  The disc is illuminated by the compact, primary emitter (i.e. the corona).  Some light bending is occurring as evident by the stepper emissivity profile in the inner part of the disc.  The emissivity profile flattens beyond about $6\rg$ at which point a highly ionised plasma, that produces the \fexxvi\ absorption, forms above the disc and extends several 10's of gravitational radii.  A ring between $13-20\rg$ could be the origin of the S~{\textsc{xvi}} that produces the absorption at $\sim2.6\keV$.  The black, dashed line is at a $45\deg$ angle and marks the line-of-sight of the observer.
Right panel:  The spectral model (solid, red curve) resulting from the geometry described in the left panel is overlaid on the 2001, $2-10\keV$ spectrum
of \pg1211.  The dashed, green curve shows the model without the resonance absorption features (effectively just the model shown in Fig.~\ref{fig:ref}).
}
\label{fig:pg1211}
\end{figure*}
%--------------------------------------------------------------------------
We now consider that this emission is exposed to resonance absorption as it emerges from the disc in a layer of highly ionised plasma that is corotating with the disc (GF11).  The plasma is thus experiencing the same blurring as the reflection spectrum (Fig.~\ref{fig:ref}).  A reasonable model is found
with continuum parameters that are similar to those described above.  The power law photon index is now $\Gamma\approx 1.84$ and the ionisation parameter of the disc is $\xi\approx 284\ergcmps$.  Iron is still over abundant to the same degree (approximately $5\times$ solar), and the disc inclination is $i\approx45\deg$.  The values of $R_{in}$,  $q_{in}$, and the break radius ($R_b$) remain comparable.  
The model also predicts a $15-50\keV$ flux of $\approx5.5\times10^{-12}\ergpscmps$, which is comparable to that reported by Patrick \et (2012) from HXD measurements. 
The $7\keV$ absorption feature can be described
as arising from \fexxvi\ that is corotating with the disc beyond the break radius where the emissivity profile flattens.  A cartoon illustrating the situation is presented in the left panel of Fig.~\ref{fig:pg1211}.    The model overlaid on the spectrum is shown in the right panel of Fig.~\ref{fig:pg1211}.  As can be seen, the feature is broad with absorption in the redwing extending down to $\sim5\keV$.
Adding another absorption line at $\sim 2.63\keV$ and blurring it in the same environment can describe the deviations from the continuum at those energies.  This particular features would originate in a ring between about $13-20\rg$ from the central black hole.  The line energy corresponds to absorption by H-like sulfur.

\section{Future observations with \astroh}
\label{sect:ah}

The exercise in Section~\ref{sect:picture} demonstrates how a resonance scattering toy model can describe the $2-10\keV$ spectrum of \pg1211, a
quasar that is often referred to as a classical example of a UFO AGN.  The large effective area of current telescopes like \xmm\ and \suzaku\ generate the required signal to detect such features, but the CCD resolution of the detectors does not typically provide the sufficient spectral resolution to distinguish various models.  Specifically, the features in the outflow models are much narrower than the broad features in the absorption model.

\begin{figure}
\rotatebox{270}
{\scalebox{0.32}{\includegraphics{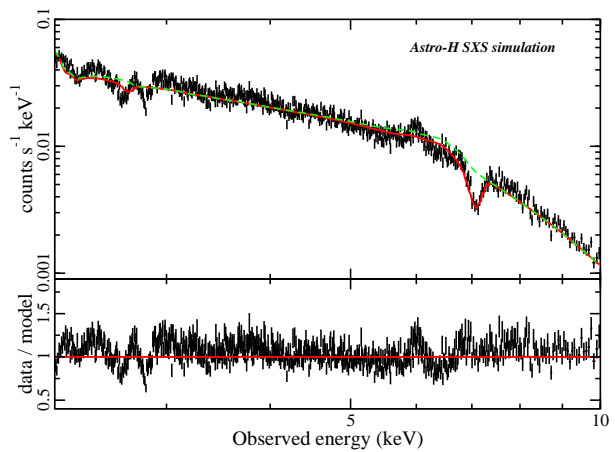}}}
\caption{Upper panel:  Simulated \astroh\ SXS data of the UFO model from Tombesi \et (2011) specifically for \pg1211.  The solid and dashed curves are the resonance absorption models, as in Fig.~\ref{fig:pg1211}, overlaid on the simulated spectrum.  Lower panel:  The residuals (data/model) highlight the difference between the sharp, narrow features in the data and the broader features in the model.
}
\label{fig:sim}
\end{figure}

\astroh\ (Takahashi \et 2012), to be launched in 2015, will be the first X-ray observatory to operate a microcalorimeter (Soft X-ray Spectrometer, SXS; Mitsuda \et 2012) that will provide high spectral resolution in the $0.3-10\keV$ band.  In Fig.~\ref{fig:sim} we present an SXS simulation of the outflow in \pg1211.  The UFO model is that of  Tombesi \et (2011) as described in Section~\ref{sect:intro} (the {\sc xstar} model was kindly provided by F. Tombesi).  The spectrum is a power law continuum modified by outflowing wind.   The turbulent velocity in the wind is $5000\kmps$.  A narrow ($\sigma=100\eV$)  \feka\ emission line is also included at $\sim6.5\keV$.
Overlaid on the simulated data is the resonance absorption model from Fig.~\ref{fig:pg1211}.    The narrow features predicted by the wind model can be distinguished from broader features.  In addition, \astroh\ will have high-energy imaging between $5-80\keV$ that will allow us to accurately constrain the broad band continuum that would differ between the blurred reflection and absorption models.
% --------------------------------------------------------------------------
% --------------------------------------------------------------------------

\section{Discussion and conclusions } 
\label{sect:dis}

The prospect of ultrafast outflows in AGN is exciting.  In most cases, the estimated mass and energy output by the system are considerable, and if
confirmed would have a prominent role in galaxy evolution.  The nature of x-ray UFOs are diverse.  In many cases, the signatures are marginally detected and transient.  In some others, the perceived outflows are inconsistent with data at other wavelengths.  Understanding the complex nature of outflows is necessary given the importance of these winds.  

GF11 proposed that at least some UFO candidates could be explained by absorption (or scattering) of resonance lines in a plasma located above and corotating with the disc.  In this way, the imprinted features would be subjected to blurring from motions of the disc.  A priori, the existence of an ionised layer  above the disc is as reasonable as a wind.  
The models are distinguishable in a number of ways.  The blurred resonance absorption features would be broader than lines from an outflow.  We would not expect to detect high velocities from discs at low inclinations, whereas an outflow might show high velocities at low inclinations unless it is equatorial.  In addition, both models predict different continuum shapes and fluxes above $10\keV$.

We show that this model can describe the features in \pg1211\ very well, an AGN that is often referred to as a classic UFO example.  
The blurring and shifting is done with velocities that are already present in the disc itself. 
The feature in \pg1211\ is rather pronounced compared to absorption features in other UFO candidates.  This could indicate that the optical depth in the line is rather high and consequently could necessitate the inclusion of an absorption edge.  However,  precise measurement of the optical depth would depend on various factors like the column density, iron abundance, and covering fraction of the absorber.  Alternatively, the absorber could be arranged in ``clumps'' of different densities and temperatures.  This could help explain the different location of the sulfur feature inferred by the model.
We fully realise that our model needs to be developed further and we are unable to make robust predictions at this time.  These are all factors that are being considered in current work.  Moreover, understanding the variability and transient behaviour is an open question for both the outflow and disc model.  If the ionised plasma is tenuous, clumpy, or occupies a small region of the disc then one may expect variability on dynamical time scales. 

In the near future, observations of UFO candidates with \astroh\ will provide the means to distinguish the proposed models.  Should our model prove correct then absorption will provide another diagnostic tool with which to probe the inner accretion flow with \astroh\ and eventually {\it Athena+}.

% --------------------------------------------------------------------------
% --------------------------------------------------------------------------

\section*{Acknowledgments}

We would like to thank Francesco Tombesi for providing his outflow model for \pg1211.  LCG would also like to thank Sherry Hurlburt and 
Dominic Walton for providing data.  We thank the referee for providing a helpful report.
ACF thanks the Royal Society for support.

%\pagebreak

%\appendix
%\section{Analysis of EPIC calibration data}
%\label{app:cal}

%\clearpage

\bsp
\label{lastpage}
\end{document}